\documentclass[prl,showpacs,twocolumn,preprintnumbers]{revtex4}
\usepackage[centertags]{amsmath}
\usepackage{amsfonts}
\usepackage{amssymb}
\usepackage{amsthm}
\usepackage{graphicx}
\usepackage{dcolumn}
\usepackage{bm}

\begin{document}

\title{The Photon Wavefunction: a covariant formulation and equivalence with QED}

\author{F. Tamburini$^\dag$}

\author{D. Vicino$^\ddag$}

\affiliation{$^\dag$ Department of Astronomy, University of Padova,
vicolo dell' Osservatorio 3, Padova, Italy.}

 \affiliation{$^\ddag$ Department of Physics, University of Padova, Via Marzolo 8,
 Padova, Italy. }

\begin{abstract}
We discuss the limits of the photon wavefunction (PWF) formalism, which is experiencing a revival in these days from the new practical applications in photonics and quantum optics.
We build a Dirac-like equation for the PWF written in
a manifestly covariant form and show that, in presence of charged matter
fields, it reproduces the standard formulation of (classical) Electrodinamics.
This shows the inconsistency of the attempts to construct
a quantum theory of interacting photons, based on the so called photon
wavefunction approach, alternative to  standard QED.
PWF formalism can then be used to provide an easier description of the propagation of free photons, when the photon number remains fixed in time.
\end{abstract}

\pacs{03.50.De, 03.65.Pm, 11.10.Ef, 12.20.m}
\maketitle

\section{Introduction}
The problem of writing a wavefunction for the photon takes its origins from
the first attempts of quantizing the electromagnetic field, since the birth
of Quantum Mechanics.
Because of the localization problem for the photon, the definition and even
the existence of a wavefunction for the photon is still controversial, as it
cannot always give a complete description of the system, like that provided by
Schr\"odinger equation for non--relativistic massive particles.
The first attempts can be found in the unpublished notes by Majorana
\cite{2 Maj}, where the quantum states of the electromagnetic field were
tentatively described by using the language of first quantization in the form
of a Dirac--like equation, obtained from the Riemann-Silberstein (RS) formulation
of Maxwell equations \cite{rie01,sil07,bo08}.
Dirac equation was formulated to describe the evolution of the relativistic
electron, a particle with non--zero rest mass, $\hbar /2$ spin, and
elementary charge $e$. Weyl equations instead describe massless neutral
spinors.
Finally Majorana extended the Dirac equation also to particles with arbitrary
spin, in a more general infinite--spin component formalism \cite{maj32}.
This approach for the quantization of the Electromagnetic field in the
first quantization language is justified by the fact that Maxwell equations
present an intrinsic mathematical structure similar to that of a quantum wave
function in relativistic theory and, conversely, the same procedure followed
by Dirac to write the relativistic equation for the electron can be used to
derive Maxwell equations.
New recent experiments with single, double and many--photon sources and also
with entangled states, where the photon number is small and remains fixed,
renewed  the interest in the wavefunction of the photon \cite{kob99, bir94,
bir96, esp98, ger01, ray05, smi06}.
This revival of interest raised some, in our opinion too
optimistic, hopes that this approach could open new perspectives for
alternative quantum descriptions of photons, even not free \cite{smi07}.
The Majorana-like equation for  the PWF can be considered as
a consistent Quantum Mechanics of a {\it free} photon only, even if modulo the
well known difficulty of its non localization \cite{bir98}.
For a system of relativistic particles in interaction  (and in
particular of interacting photons) Quantum Mechanics cannot be used since
it implies action-at-distance forces, incompatible with Relativity.
Quantum Field Theories are, in fact, mandatory.

In this letter, after rewriting the Majorana-like equation of the photon
wavefunction in a manifest covariant form, we discuss the lagrangian that
reproduces this equation, to be used as the starting point for getting
a Quantum Field Theory for (interacting) photons. Then we show that this
lagrangian is equivalent to that of classical Electrodynamics so that after
quantization it leads to standard QED.

\section{PWF and covariant formulation} 
Following Majorana formulation, without loosing in generality for a particular
choice of helicity state, or of multiplicative constants, one defines the 
Riemann--Silberstein vector
\begin{equation}
\label{F} \mathbf{F}=\frac{\textbf{E}}{c} \pm i\textbf{B}
\end{equation}
and Maxwell Equations in the vacuum become
\begin{eqnarray}
\label{Maxwell}
\nabla \cdot \mathbf{F}=0, \quad
i\nabla\times \mathbf{F}=\pm \frac{1}{c}\frac{\partial
\mathbf{F}}{\partial t}
\end{eqnarray}
By using the correspondence principle,  $\mathbf{p} \leftrightarrow
\hat {\textbf{p}} \equiv -i \hbar \nabla$ ($i=1, 2, 3$), \textbf{F}
here represents the wavefunction of the photon, leading to the
wave equation $\mp \frac{i \hbar}{c}\frac{\partial}{\partial
t}\mathbf{F}+i\mathbf{\hat{p}}
 \times \mathbf{F}=0$, while
$\nabla \cdot \mathbf{F}=0$ is the transversality of the fields with respect
to the propagation direction, namely $\hat{\textbf{p}}\cdot \mathbf{F}=0$.
By introducing the $3 \times 3$ complex matrices
\begin{equation} \nonumber
    \hat{s}_x=\left(
            \begin{array}{ccc}
            0&0&0\\
            0&0&-i\\
            0&i&0
            \end{array}
            \right),
    \hat{s}_y=\left(
            \begin{array}{ccc}
            0&0&i\\
            0&0&0\\
            -i&0&0
            \end{array}
            \right),
    \hat{s}_z=\left(
            \begin{array}{ccc}
            0&-i&0\\
            i&0&0\\
            0&0&0
            \end{array}
            \right)
\end{equation}
\normalsize
one obtains a Dirac--like equation
\begin{equation} \label {Dirac}
i \hbar \frac{\partial}{\partial t} \mathbf{F}= \hat{H}
\mathbf{F},
\end{equation}
with  $\hat{H}=\pm c\,\hat{\mathbf{s}}\cdot \hat{\textbf{p}}$
and $\hat{\textbf{s}}=(\hat{s}_x, \hat{s}_y, \hat{s}_z)$  \cite{2 Maj, bir96,kob99}.
This Hamiltonian has eigenvalues $ \pm cp,0$. The eigenvalue $0$ is
forbidden by the transversality condition.

There has been a debate in the literature on the interpretation of
the \textit{negative energy state} with eigenvalue $- cp$. The
interpretation of the states with eigenvalues $\pm cp$ as states of
positive energy and helicity $\pm 1$ is discussed in  \cite{bir96}.
A simple way to understand this point is the following: since in our
case the observables of Energy and Helicity commute, we can
interpret the generator of translations in time, $ \hat H $ in
(\ref{Dirac}) (after diagonalization) as the product  of the
Hamiltonian $ H=c p$ with the helicity operator $\lambda$. The wave
equation now has energy eigenvalues  always positive and the negative eigenvalue of $\hat H$ is due to the negative helicity value.
As for the trasversality condition $\nabla \cdot \mathbf{F}=0$, let us recall
 that it is at the origin of the non localization of the PWF.
Indeed the Hilbert space where $F(x)$ lives, is the space of modulo,
square functions $\phi(x)$ that satisfy the condition $\nabla \cdot
\mathbf{\phi}=0$ and therefore, the \textit{position operator} $ X\phi(x)
= x\phi(x)$ is not an operator of this space, since $x\phi(x)$ does
not satisfy the transversality condition.

The first step to a covariant formulation for \textbf{F} is the isophormism
between the algebras of the group $SL(2 \mathbb{C})$, of unimodular
$2\times2$ matrices in the complex field and the (proper orthocronus)
Lorentz group, $SO(1,3)$, of  $4\times4$ (pseudo-)orthogonal real matrices
that leave invariant  the Minkowsky metrics $\eta^{\mu\nu}= diag (1,-1,-1,-1)$.
Since the two algebras are isomorphic, the two groups satisfy a local
isomorphism  that, extended to a global one, becomes a  $2 \rightarrow 1$
homomorphism. Consider the matrix
\begin{equation}\label{transf}
\overline{x}= \left(  \begin{array}{cc}
x_0+x_3& x_1-ix_2\\
x_1+ix_2& x_0-x_3
\end{array}
\right)
\end{equation}
built with the  space-time coordinates,  $x_0,x_1,x_2,x_3$ and transform
$\overline{x}$ by an  $SL(2 \mathbb{C})$ trasformation as
$\overline{x'}=A^{-1}\overline{x} A$,
where $A \in SL(2 \mathbb{C})$, so that $det \overline{x'} = det
\overline{x}$.
Since $det \overline{x} = x_0^2-|\mathbf{x}|^2
= \eta^{\mu \nu}x_{\mu}x_{\nu}$, the transformation 
 leaves invariant the four dimensional interval $s^2=\eta^{\mu
\nu}x_{\mu}x_{\nu}$ and therefore induce a Lorentz transformation $\Lambda$
on $x^{\mu}$, but to both $\pm A$ correspond the same Lorentz transformation
(i.e. the homomorphism is  $2 \rightarrow 1$ ).
The group $SL(2 \mathbb{C})$ has two inequivalent, fundamental representations
called $\left(\frac{1}{2},0 \right)$ and $\left(0, \frac{1}{2}\right)$ that
can be also considered as two spinorial representations of $ SO(1,3)$ (chiral
and antichiral Weyl spinors).

The elements of the vector space on which these representations
operate are $ \phi_{\alpha}(x), ~ (\alpha=1,2)$ for the
$\left(\frac{1}{2},0\right)$ and     $ \overline{\phi}^{\dot
{\alpha}} \equiv \varepsilon^{\dot{\alpha}\alpha} \phi_{\alpha}^*, ~
(\dot{\alpha}, \alpha=1,2) $for  $\left(0, \frac{1}{2}\right)$, such
that $\phi'_{\alpha}(x')={A_{\alpha}}^{\beta} \phi_{\beta}(x)$ and
${\overline{\phi}'}^{\dot{\alpha}}(x') ={(A^{\dag \,
-1})^{\dot{\alpha}}}_{\dot{\beta}} \overline{\phi}^{\do {\beta}}(x)$
where  $A, A^{\dag \, -1}\in SL(2 \mathbb{C})$

Tensorial products of the fundamental representations of the Lorentz  group
 give rise to higher dimensional representations, divided in two main classes:
 \textit{Tensorial} representations that derive from the product of even
times the fundamental representations ($\frac12,0$) and/or ($0,\frac12$),
and \textit{Spinorial} representations coming from the product of odd ones.
The simplest tensorial representations are
$\psi_{\alpha}^{\dot{\beta}}\equiv
\phi_{\alpha}\otimes\overline{\chi}^{\dot{\beta}}$,
  $\psi_{\alpha \beta}\equiv \phi_{\alpha}\otimes\chi_{\beta}$,
$\psi_{\dot{\alpha}\dot{\beta}}\equiv
\overline{\phi}_{\dot{\alpha}}\otimes\overline{\chi}_{\cdot{\beta}}$,
with $\phi_{\alpha}, \, \chi_{\beta} \in \left(\frac{1}{2},0
\right)$, $\overline{\phi}_{\dot{\alpha}}, \,
\overline{\chi}_{\dot{\beta}} \in \left(0,\frac{1}{2} \right)$.
Written in its symmetric and antisymmetric parts, $\psi_{\alpha \beta}
=\psi_{[\alpha \beta]} +\psi_{(\alpha \beta)}$, then, $\psi_{[\alpha
\beta]}=\lambda \varepsilon_{\alpha\beta}$, has the only degree of freedom
$\lambda$, where $\varepsilon^{\alpha \beta}$ is the $2 \times 2$ completely
antisymmetric tensor.
The symmetric part has instead $3$ independent components.
A similar decomposition holds for $\psi_{\dot{\alpha}\dot{\beta}}$.
The constant tensors  $\varepsilon_{\alpha \beta}$, $\varepsilon_{\dot
{\alpha}\dot{\beta}}$ and their inverses  $\varepsilon^{\alpha \beta}$,
$\varepsilon^{\dot{\alpha}\dot{\beta}}$ can be used to rise and lower the
spinorial indices.
Let us define the two by two matrices ${({\sigma}^\mu)_{\gamma}}^{\dot{\beta}}$
and ${(\overline{\sigma}^\mu)_{\dot{\alpha}}}^{\gamma}$ where
$\overline{\sigma}^0 ={\sigma}^0 = 1 $, $\overline{\sigma}^i =  - {\sigma}^
i $ and  ${\sigma}^i$ are the Pauli matrices. The tensorial representations
$\psi_{\alpha}^{\dot{\beta}}$, $\psi_{\alpha \beta}$ and
$\psi_{\dot{\alpha}\dot{\beta}}$ can be expressed in terms of these matrices,
\begin{eqnarray}
&&\psi_{\alpha}^{\dot{\beta}}= {({\sigma}^\mu)_{\alpha}}^{\dot{\beta}}\psi_{\mu}
\\
&&\psi_{(\alpha\beta)} =(\sigma^{[\mu}\overline{\sigma}^{\nu
]})_{(\alpha \beta)}\psi_{[\mu \nu]}
\\
&&\psi_{(\dot{\alpha}\dot{\beta})} = (\overline{\sigma}^{[\mu}\sigma^
{\nu ]})_{(\dot{\alpha}\dot{\beta})}\psi_{[\mu \nu]}
\end{eqnarray}
but the first of these equations, that defines a four-vector $\psi_{\mu}$. Similarly,
\begin{equation}
\psi_{[\mu \nu]}^{\pm} = \psi_{[\mu\nu]} \pm i\frac{1}{2}\varepsilon_{\mu\nu\rho
\sigma}\psi^{\rho\sigma}.
\end{equation}
Both $\psi_{\mu\nu}^+$ and $\psi_{\mu\nu}^-$ have $3$ independent components
and
${\psi^+_{\mu\nu}}^D=\psi^+_{\mu\nu}, \quad
{\psi^-_{\mu\nu}}^D=-\psi^-_{\mu\nu}$, where ${\psi_{\mu\nu}}^D=i\frac{1}{2}
\varepsilon_{\mu\nu\rho\sigma}\psi^{\rho\sigma}$ is the dual tensor of
$\psi_{\mu\nu}$.

One can verify easly that $\psi^{-}$ and $\psi^{+}$ do not contribute to
$\psi_{\alpha \beta}$ and to $\psi_{\dot{\alpha}\dot{\beta}}$ respectively
so that
\begin{eqnarray}
&&\psi_{(\alpha\beta)} =(\sigma^{[\mu}\overline{\sigma}^{\nu ]}\psi_{[\mu \nu)]}
^{+})_{\alpha\beta}
\\
&&\psi_{(\dot{\alpha}\dot{\beta})} = (\overline{\sigma}^{[\mu}\sigma^
{\nu ]}\psi_{[\mu \nu]}^{-})_{\dot{\alpha}\dot{
\beta}}.
\end{eqnarray}
Therefore $\psi_{(\alpha\beta)}$ describes the self--dual part,
$\psi_{[\mu\nu]}^{+}$ and
$\psi_{(\dot{\alpha}\dot{\beta})}$ the antiself--dual part,$\psi_{[\mu\nu]}^{-}
$, of a six-components double--antisymmetric tensor $\psi
_{[\mu\nu]} \in \left(\frac{1}{2}\,\frac{1}{2},0\right)\oplus
\left(0,\frac{1}{2}\,\frac{1}{2}\right)$.

The Faraday electromagnetic tensor $F_{\mu\nu}$ such that $F^{0i} =
- \frac {1}{c}E_{i}$ and $ F^{ij}= - \varepsilon^{ijk} B_{k}$ is
a double antisymmetric tensor and its self-dual and anti
self-dual parts $\left( F^+_{\mu \nu}, \, F^-_{[\mu \nu]} \right)$
can be written in covariant spinor notation
\begin{eqnarray}
&&F_{(\alpha\beta)}=
(\sigma^{[\mu}\overline{\sigma}^{\nu]}{F^+_{[\mu \nu]}} )_
{\alpha\beta}
\\
&&\overline{F}_{(\dot{\alpha}\dot{\beta})}=(\overline{\sigma}^{[\mu}\sigma^
{\nu]}{F^-_{[\mu\nu]}} )_{\dot{\alpha}\dot{\beta}}
\end{eqnarray}
Where $F_{(\alpha\beta)} \in
\left(\frac{1}{2}\,\frac{1}{2},0\right)$ and 
$\overline{F}_{(\dot{\alpha}\dot{\beta})} \in
\left(0,\frac{1}{2}\,\frac{1}{2}\right)$, but
\begin{eqnarray}
F_{(\alpha\beta)}&\propto&\left((\sigma^{[0}\overline{\sigma}^{i]})F_{0i}^+\right)_{\alpha
\beta} \equiv \left(\overline{\sigma}^{i} F_{i}^{+}\right)_{\alpha
\beta},
\\
\overline{F}_{(\dot{\alpha}\dot{\beta})}&\propto&\left((\overline{\sigma}^{[0}\sigma^{i]})F_{0i}^-\right)_{\dot{\alpha}
\dot{\beta}} \equiv \left(\sigma^{i} F_{i}^{-}\right)_{\dot{\alpha}
\dot{\beta}}
\end{eqnarray}
and
\begin{equation}
F_{i}^{\pm} = \frac{E_{i}}{c} \pm i B_{i}
\end{equation}
so that $F_{(\alpha\beta)} $ and $ \overline{F}_{(\dot{\alpha}\dot{\beta})}$ are
the positive  and negative helicity wavefunctions of the photon in covariant
notations.

To write the Dirac-like equation for the photon wave
function, consider
\begin{eqnarray}
\label{eq_onda_cov_+}
{(\overline{\sigma}^{\mu}\partial_{\mu})_{\dot{\alpha}}}^{\beta} F_{(\beta
\alpha)}&=&
(\overline{\sigma}^{\mu}\partial_{\mu}\sigma^{\nu}\overline{\sigma}^{\lambda})
_{\dot{\alpha}\alpha}F^+_{\nu\lambda}=0
\end{eqnarray}
and its complex conjugate
\begin{eqnarray}
\label{eq_onda_cov_-}
{(\sigma^{\mu}\partial_{\mu})_{\alpha}}^{\dot{\beta}}\overline{F}_{(\dot{
\beta}\dot{\alpha})}&=&
(\sigma^{\mu}\partial_{\mu}\overline{\sigma}^{\nu}\sigma^{\lambda})
_{\alpha \dot{ \alpha}}F^-_{\nu\lambda}=0.
\end{eqnarray}
When eqn.(\ref{eq_onda_cov_+}) is saturated with $(\sigma^{\tau})^{\alpha \dot{
\alpha}}$ one gets, for $\tau =0 $, the first equation in (\ref{Maxwell})
and, for $\tau = i$, the second equation in  (\ref{Maxwell}) that is equation
(\ref{Dirac}). The same results are obtained saturating 
(\ref{eq_onda_cov_-}) with $(\overline{\sigma}^\tau)^{\dot{\alpha}\alpha}$.

Equation (\ref{Dirac}), toghether with the transversality condition
$ \nabla \cdot \mathbf{F}=0 $, is equivalent to free Maxwell equations.
This leads to speculate that this approach could be taken as the starting 
point for a new quantum description of, even not free, photons. 
However when it is rewritten in covariant form,
(\ref{eq_onda_cov_+}), it becomes completely clear that it describes just free
Maxwell equations in a different notation. That leaves little room to the 
speculations previously mentioned.

Let us add some further considerations to stress this point ever more.
As noted in the introduction, a relativistic quantum theory with
interactions must be necessarly a local  QFT. The recipe to write
the (free) classical field lagrangian density, to be quantized, is
to look at the classic action that yields  the Schroedinger equation
of the Quantum Mechanics of the single particle (eventually
supplemented with local interaction terms)
 and quantize this classical action according to the canonical rules. This
procedure has been named, quite improperly, second quantization.

For instance the lagrangian density of a free electron, derived from the
Dirac equation, is  $\mathcal{L}_e=\overline{\Psi}\left(i \gamma^{\mu}
\partial_{\mu}-m I_4 \right)\Psi$,
which is invariant under the global gauge transformation $\Psi \rightarrow
e^{i \lambda}\Psi$, where $\lambda$ is the global gauge parameter.
To extend this trasformation to a local one with gauge parameter
$\lambda(x)$ one must introduce a gauge field $A_{\mu}$ that trasforms as
$ A_{\mu} \rightarrow  A_{\mu} + \partial_{\mu}\lambda$ and the lagrangian
density becomes
\begin{equation}\label{Lagrangiana_el_v}
\mathcal{L}_e= \overline{\Psi}\left(i \gamma^{\mu}
(\partial_{\mu}-ie A_{\mu})-mI_4\right)\Psi.
\end{equation}

Now we have to search for a lagrangian density that gives rise to
(\ref{eq_onda_cov_+})
as its Eulero-Lagrange field equations. Since
(\ref{eq_onda_cov_+}) transform covariantly as a four-vector one needs a
four-vector,
let say written in spinor notations, $A^{\alpha \dot{\alpha}}= A_{\mu}
(\sigma^{\mu})^{\alpha\dot{\alpha}}$. A lagrangian density that reproduces
(\ref{eq_onda_cov_+}) by varying ${A}$ is
$\mathcal{L}= a (A\overline{\sigma}^{\mu}\partial_{\mu})^{\alpha
\beta}F_{\alpha \beta}$
where $a$ is a normalization constant.
Notice that this lagrangian is invariant under the gauge trasformation
$ A_{\mu} \rightarrow  A_{\mu} + \partial_{\mu}\lambda$.

If one adds to the lagrangian  $\mathcal{L}$ 
the Dirac lagrangian $\mathcal{L}_{e}$ it is right to identify, as
anticipated with the notations, the real four-vector in ${A}$ with
the $A_{\mu}$ in (\ref{Lagrangiana_el_v}) since with this
identification the field equations for $A_{\mu}$ yield correctly the
current term,  $j^{\mu}= e \overline{\Psi}\gamma^{\mu}\Psi$ in the
right hand side of the first group of Maxwell equations (or
equivalently a non linear term $j^{i}$ in the r.h.s. of
(\ref{Dirac}) and a non linear term $j^{0}$ in the r.h.s. of $
\nabla \cdot \mathbf{F}=0 $). However the field equations obtained
varying $F_{(\alpha\beta)}$ are $(\partial_{[\mu}A_{\nu]})^{+} = 0$
that imply $A_{\mu}=0$ modulo a gauge transformation.  But
$\mathcal{L}$ has a serious drawback: it is not real.

By adding to $\mathcal{L}$ 
the complex conjugate counterpart, the  action becomes
\begin{eqnarray}\label{Lagrangiana2}
 I &=& \int \frac{a}{2}[({A}\overline{\sigma}^{\mu}\partial_{
\mu})^{\alpha\beta}F_{\alpha \beta} +  (\overline{A}{\sigma}^{\mu}
\partial_{\mu})^{\dot \alpha\dot \beta}\overline{F}_{\dot\alpha \dot \beta}] +
 \int \mathcal{L}_e \nonumber
\\
&=& \int a ( A_{\mu} \partial_{\nu}F^{\mu\nu}) +\int \mathcal{L}_e
\end{eqnarray}
where  $\overline{A} =A_\lambda \overline\sigma^\lambda$. Unfortunatly
(\ref{Lagrangiana2}) reproduces only the first group of Maxwell equations
$ \partial^{\mu}F_{\mu\nu} = e j_{\nu}$ (with $a = 3/4 $).

A possible cure of this desease could be to allows   $A$
 and $\overline{A}$ to become complex that is
 ${A}=(A_{\lambda}  +iB_{\lambda})\sigma^\lambda$, $\overline{A}
 =(A_{\lambda} - iB_{\lambda}) \overline\sigma^\lambda$ ($A_{\mu}$ and
$B_{\mu}$
real). Now (\ref{eq_onda_cov_+}) (with current) are reproduced correctly but
there is a unacceptable doubling of degree  of freedom. Indeed now the action
(\ref{Lagrangiana2}) becomes
\begin{equation}\label{Lagrangiana3}
I = \int ( A_{\mu} \partial_{\nu}F^{\mu\nu}  + B_{\mu}\varepsilon^{\mu\nu
\rho\sigma}\partial_{\nu}F_{\rho\sigma} + \mathcal{L}_e)
\end{equation}
The second group of Maxwell equations $ \varepsilon^{\mu\nu\rho\sigma}
\partial{\nu}F_{\rho\sigma} =0$ implies $ F_{\mu\nu} = \frac12 ( \partial_{\mu}
\tilde A _{\nu} - \partial_{\nu}\tilde A _{\mu})$.  When this algebraic
equation
is used to remove $F_{\mu\nu} $ in (\ref{Lagrangiana3}) the $B_{\mu}$ field
drops out and the action becomes
$I = \int ( \partial_{[\mu}A_{\nu]}  \partial^{[\mu} \tilde A^{\nu]}  +
\mathcal{L}_e)$
Even worse, if one defines ${A^{(\pm)}}_{\mu} = \frac{1}{2}(A_{\mu}\pm \tilde A_
{\mu}) $, then 
\begin{equation}
\label{Lagrangiana5} I = \int (
\partial_{[\mu}A^{(+)}_{\nu]}\partial^{[\mu} A^{\nu](+)} -
 \partial_{[\mu}A^{(-)}_{\nu]}\partial^{[\mu} A^{\nu](-)}
+ \mathcal{L}_e )
\end{equation}
so that, after quantization, one of the two ``photons'' described by the
gauge fields ${A^{\pm}}_
{\mu}$  has negative metric and therefore the action
(\ref{Lagrangiana5}) is inconsistent.

The only consistent way to cure these problems is to add to the lagrangian
density in (\ref{Lagrangiana2}) the term
\begin{equation}
\mathcal{L}_{0} = - \frac{3}{32}
(F^{\alpha\beta}F_{\beta\alpha} + \overline{F}^{\alpha\beta}\overline{F}_{
\beta\alpha}) =  \frac{1}{2}F^{\mu \nu}F_{\mu \nu},
\end{equation}
to obtain
\begin{eqnarray}
I = \int\mathcal{L}_{tot}= -\int \partial^{[\mu]}A^{ \nu]}F_{\mu\nu}
 -\int \frac{1}{2}F^{\mu \nu}F_{\mu \nu}
\\
+  \int \overline{\Psi}\left(i \gamma^{\mu}
(\partial_{\mu}-ieA_{\mu})    -m I_4 \right)\Psi  \nonumber
\end{eqnarray}
The field equation for $F_{\mu\nu}$ identifies  $F_{\mu\nu}$ with $\partial_
{[\mu}A_{\nu]}$:  $F_{\mu\nu} = - 1/2(\partial_{\mu}A_{\nu}-\partial_{\nu}A_
{\mu})\equiv \partial_{[\mu}A_{\nu]} $.
Since this is an algebraic equation it can be used to replace  $F_{\mu\nu}$ in the lagrangian with the following result,
\begin{eqnarray}
\label{action2}
I &=& \frac{1}{2} \int  \partial_{[\mu}A_{\nu]}\partial^{[\mu}A^{\nu]}
\\
 &+& \int \overline{\Psi}\left( i \gamma^{\mu}
(\partial_{\mu}-ieA_{\mu}) - m I_4 \right)\Psi \nonumber
\end{eqnarray}
eqn. (\ref{action2}) is the standard action of classical electrodynamics and after
quantization it gives rise standard QED.
In the presence of charged matter fields, this formulation reproduces only the standard formulation of (classical) Electrodinamics.
This clearly shows the inconsistency of the attempts to construct
a quantum theory of interacting photons, based on the so called photon
wavefunction approach, alternative to  standard QED.
There is a perfect correspondence between PWF and QED only when
photons are free, non interacting and when the photon number remains
constant during the evolution of the field, with the problems of the
photon localization.

\section{Discussion and Conclusions}
In this note we have presented and discussed the equivalence of
the PWF formalism with that of standard Quantum Electrodynamics.
PWF can only describe scenarios where the photons are free,
non interacting and maintain a constant number during their evolution.
No absorption and/or emission of photons can be directly described by the formalism of PWF. 
The equivalence is set by a manifestly covariant
version of the so called PWF equation. Moreover, on the basis
of this covariant formulation we have motivated the statement
that the photon wave function approach, at the second
quantization level, cannot give anything else than the standard Quantum
ElectroDynamics.

The correspondence set between PWF and QED for free photons
is useful to shed some light in the Orbital Angular Momentum (OAM) of the photon and the PWF formalism \cite{arl98,mai01,MT02,oneil02}. 
OAM of light is deeply connected with the vorticity of the E-M
field and with the creation of optical vortices.
Recently the RS vector, which is the basis with which the PWF is
built, was used to describe the E-M field vorticity \cite{ber04,ber04b,rad04,bir06}.
Riemann Silberstein vortices are defined by
\begin{equation}
{\bm F}({\bm r}, t) \cdot {\bm F}({\bm r}, t) =0.
\end{equation}
The loci of points satisfying this condition are lines in space, the phase of the field is
singular surrounded by zones where the phase gradient vector is circulating.
Laguerre--Gaussian beams are particular cases in which the field has spatial
symmetry and the RS vortex lines are stationary.
Anyway exact solutions of electromagnetic waves carrying angular momentum
have been recently described in by using the momentum representation and then
were cast in terms of PWF in the RS formalism \cite{bir06} that,
by using the correspondence here discussed, is simply equivalent to the description
obtained with QED by quantizing the field in paraxial approximation that,
at the single photon level, represents the probability
amplitude of finding a photon in a certain eigenstate of momentum, helicity and OAM \cite{cal06}, that is clearly not an intrinsic property of the photon\footnote{By definition, the intrinsic properties of a particle are those that do not depend on the choice of a reference frame, i.e the rest mass, the electric charge and spin.}. 
The equivalence of the two formulations can be easily set by expressing the PWF in terms of the photon annihilation and creation operators, using the vector potential $A$ in the RS vector, but this
goes beyond the purpose of this work.

\acknowledgments
We would like to thank Mario Tonin for the invaluable help in this work.
FT also acknowledges the financial support from the CARIPARO foundation.

\end{document}